\documentclass[preprint,12pt]{aastex}
\usepackage{natbib}
\newcommand {\apgt} {\ {\raise-.5ex\hbox{$\buildrel>\over\sim$}}\ }
\newcommand {\aplt} {\ {\raise-.5ex\hbox{$\buildrel<\over\sim$}}\ } 
\shorttitle{Inverse $T$ - $\beta$ Correlation due to Noise}
\shortauthors{Shetty et al.}

\begin{document}
\title{The Effect of Noise on the Dust Temperature - Spectral Index
Correlation} \author{Rahul Shetty\altaffilmark{1,2}, Jens
Kauffmann\altaffilmark{1,2}, Scott Schnee\altaffilmark{3}, Alyssa
A. Goodman\altaffilmark{1,2}}

\altaffiltext{1}{Harvard-Smithsonian Center for Astrophysics, 60
Garden Street, Cambridge, MA 02138} 
\altaffiltext{2}{Initiative for
Innovative Computing, Harvard University, 60 Oxford Street, Cambridge,
MA, 02138} 
\altaffiltext{3}{Division of Physics, Mathematics, and
Astronomy, California Institute of Technology, 770 South Wilson
Avenue, Pasadena CA 91125} 

\email{rshetty@cfa.harvard.edu}

\begin{abstract}
We investigate how uncertainties in flux measurements affect the
results from modified blackbody SED fits.  We show that an inverse
correlation between the dust temperature $T$ and spectral index
$\beta$ naturally arises from least squares fits due to the
uncertainties, even for sources with a single $T$ and $\beta$.
Fitting SEDs to noisy fluxes solely in the Rayleigh-Jeans regime
produces unreliable $T$ and $\beta$ estimates.  Thus, for long
wavelength observations ($\lambda$\apgt200 \micron), or for warm
sources ($T$\apgt 60 K), it becomes difficult to distinguish sources
with different temperatures.  We assess the role of noise in recent
observational results that indicate an inverse and continuously
varying $T-\beta$ relation.  Though an inverse and continuous
$T-\beta$ correlation may be a physical property of dust in the ISM,
we find that the observed inverse correlation may be primarily due to
noise.

\end{abstract}


\section{Introduction\label{introsec}}

Recent advances in infrared and sub-millimeter observations have
enabled detailed investigation of the properties of dust in a wide
range of environments.  Both ground and space based observatories,
such as {\it SCUBA}, {\it Bolocam}, {\it MAMBO}, {\it Spitzer}, and
{\it ISO}, as well as the balloon borne experiment {\it PRONAOS}, have
revealed much about the nature of dust emission.  The near-future
observatories {\it Herschel} and {\it Planck} will also be capable of
detecting dust in a variety of environments.  One conclusion drawn
from spectral energy distribution (SED) fits to flux measurements by
{\it PRONAOS} is that the dust emissivity spectral index $\beta$ has
an inverse correlation with the dust temperature $T$, and that this
correlation is well described as a hyperbola \citep{Dupacetal03}.
\citet{Desertetal08}, using longer wavelengths fluxes of galactic
sources from the Archeops experiment, and \citet{Yang&Phillips07},
through observations of luminous infrared galaxies (LIRGS), also find
a similar anti-correlation.

In this short paper, we present an investigation of the effect of
noise on the $T$ and $\beta$ estimates from SED fits to flux
measurements.  In recent work, we focused on starless cores, with
$T$\aplt20 K, and showed that SED fits may result in inversely
correlated and erroneous $T$ and $\beta$ estimates, due to line of
sight temperature variations, noise, or both (Shetty et al., ApJ
submitted, hereafter Paper I).  Here, we consider warmer sources.  We
investigate how uncertainties in the fluxes, coupled with the
$T-\beta$ degeneracy that arises from the functional form of the
modified blackbody spectrum describing dust emission
\citep{Blainetal03,Sajinaetal06}, lead to a spurious inverse $T-\beta$
correlation from SED fits.

\section{Modified Blackbody Assumption \label{methsec}}
The common assumption about the SED due to dust is that it is a
blackbody modified by a power law in the frequency
\citep{Hildebrand83}.  For optically thin dust emission, the flux
density $S_\nu$ takes the form
\begin{equation}
S_\nu=\Omega B_\nu(T) \kappa_0 \left( \frac{\nu}{\nu_0} \right)^\beta N.
\label{fd}
\end{equation}
Here, $\Omega$ is the solid angle of the observing beam, $B_\nu (T)$
is the Planck function, and $N$ is the dust column density.  The term
$\kappa_0 (\nu/\nu_0)^\beta$ is the frequency dependent opacity of the
emitting dust.  Dust in the interstellar medium is usually
characterized by 1 \aplt $\beta$\aplt 2 \citep[e.g.][and references
therein]{Draine&Lee84,Mathis90}.  Equation (\ref{fd}) only accurately
characterizes sources with a single $T$ (isothermal) and a single
$\beta$.

With 4 or more fluxes, a direct fit of equation (\ref{fd}) to the
fluxes can be employed to estimate three parameters: $\beta$, $T$, and
$\kappa_0 N$ (Paper I, and references therein).  As discussed in Paper
I, the fitting results are sensitive to noise and temperature
variations along the line of sight.  A single temperature fit to
fluxes near the peak of the SED may result in erroneous $T$ and
$\beta$ estimates, due to line of sight temperature variations.  For
fluxes in the Rayleigh-Jeans (R-J) tail, the best fit $T$ may provide
an accurate estimate for the ``column temperature,'' or density
weighted temperature.  However, in the R-J tail, the fits are very
sensitive to noise uncertainties.  In this work, we consider
isothermal sources, focusing on the effect of noise on $T$ and $\beta$
estimates from SED fits.

\section{Effect of Noise on SED Fits \label{sedfitsec}}

In order to assess the effect of noise on the SED fits, we employ
simple Monte Carlo experiments.  We compute the fluxes at various
wavelengths from sources with given temperatures and spectral indices,
using equation (\ref{fd}).  To incorporate the effect of noise, we
then multiply each flux by a factor 1+$\epsilon$, where $\epsilon$ is
a random value drawn from a Gaussian distribution with a chosen
dispersion $\sigma$ (and mean of 0).  We then perform a minimized
Chi-squared fit of equation (\ref{fd}) to the noisy fluxes, and
compare the resulting $\beta$ and $T$ estimates to those of the
sources.

\begin{figure}
\plotone{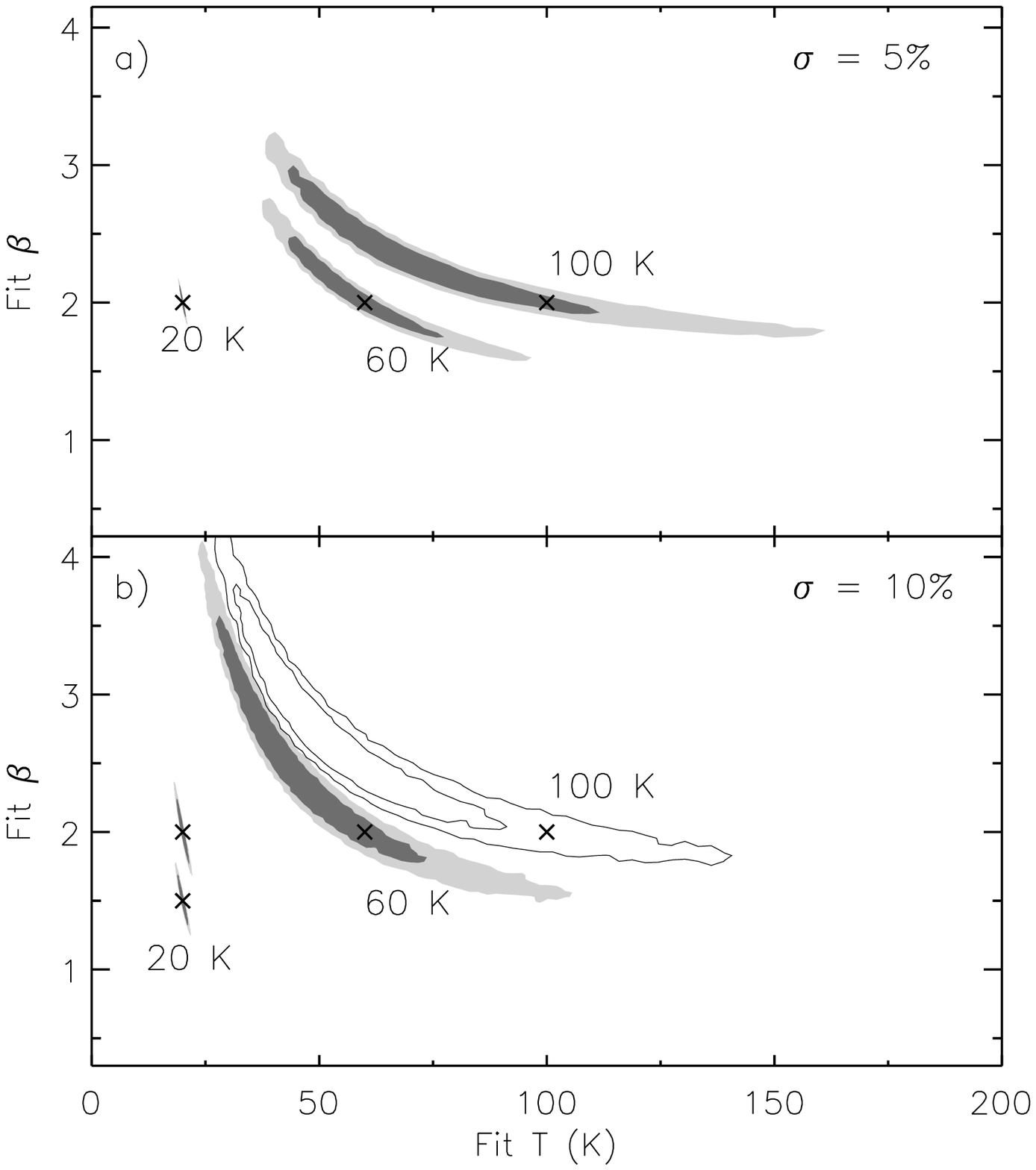}
\caption{The 75\% and 50\% probability contours for the best fit $T$
and $\beta$ to noisy fluxes, in the wavelength range 100-600 \micron\
(sampled at 10 \micron\ increments), from 20 K, 60 K, and 100 K
isothermal sources with $\beta=2$ (marked by crosses).  Gaussian
distributed noise is included in each flux, with (a) $\sigma$=5\% and
(b) $\sigma$=10\%.  In (b), best fit $T$ and $\beta$ from a 20 K
source with $\beta$=1.5 is also shown.}
\label{range}
\end{figure}

Figure \ref{range} shows the 75\% and 50\% probability contours of the
best fit $T$ and $\beta$ for three isothermal sources, with $T$=20,
60, and 100 K, and with $\beta$=2.  To obtain good statistics, we
perform fits to numerous sets of data.  The fluxes are sampled at 10
\micron\ intervals, in the range $\lambda \in$ 100-600 \micron.  The
noise levels of the fluxes are 5\% and 10\%.  For a 5\% error in the
fluxes, the mean (or median) of the $T$ and $\beta$ fits provide a
good estimate of the source parameters for the 20 K source. Though an
inverse $T-\beta$ relationship is apparent, the spread about the
source parameters is very small.  However, for the warmer sources, the
fits show a clear inverse $T-\beta$ relationship; the mean or median
$T$ and $\beta$ also do not correspond to the source values.  We note
that weighting the fluxes by signal-to-noise, or fitting in
log($\lambda$)-log($S_\nu$) space, can improve the fits, but an
inverse $T-\beta$ relationship still emerges.  In this work all the
fluxes are weighted equally, in order for direct comparison with
recent investigations in $\S$\ref{compsec}.

Also shown in Figure \ref{range}(b) are fits to fluxes from a 20 K
source with $\beta$=1.5.  These fits illustrate that sources with
different values of the spectral index, but with identical
temperatures, would give vertically offset, but otherwise similar,
points in a $T-\beta$ plane.

\begin{figure}
\plotone{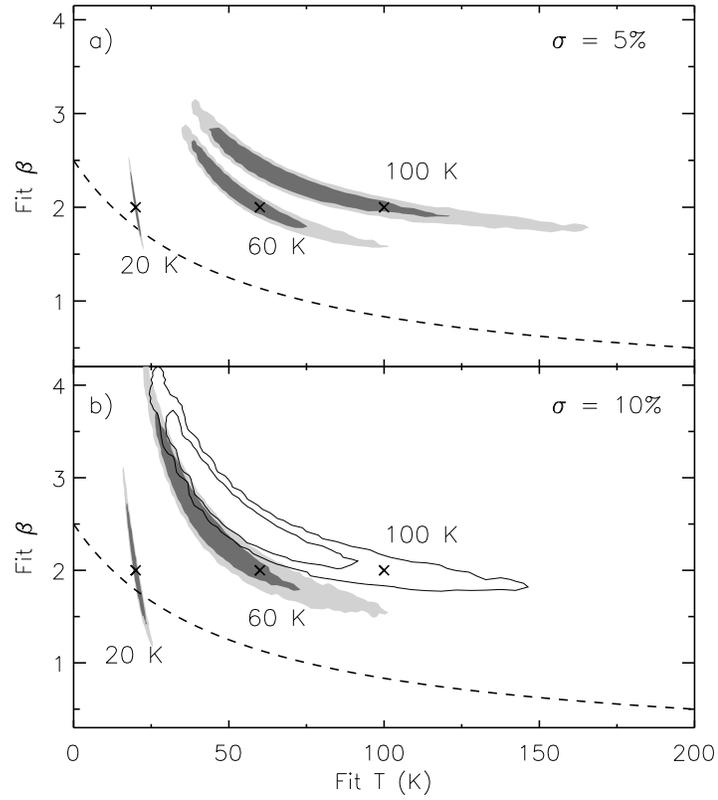}
\caption{The 75\% and 50\% probability contours for the best fit $T$
and $\beta$ to noisy fluxes, with (a) $\sigma$=5\% and (b)
$\sigma$=10\%.  The wavelengths of the sampled fluxes are
$\lambda$=100, 200, 260, 360, and 580 \micron, from three isothermal
sources (as in Fig. \ref{range}).  The dashed line is the best fit
hyperbola from \citet{Dupacetal03}.}
\label{few}
\end{figure}

In practice, only a few fluxes are typically available from
observations for fitting an SED.  To simulate this more realistic
scenario, we consider fits to sparsely sampled fluxes.  Figure
\ref{few} shows the 75\% and 50\% likelihood contours of the best fit
$T$ and $\beta$ from fits to five fluxes, with $\lambda$=100, 200,
260, 360, and 580 \micron.  In comparison with Figure \ref{range}, the
spread in both $\beta$ and $T$ estimates is larger, especially for
fits to fluxes from the 20 K source.  The fits from the warmer 60 K
and 100K sources are less distinguishable from each other, compared
with the analogous fits incorporating many more fluxes shown in Figure
\ref{range}.  Further, fits to fluxes from the warmer sources are less
likely to recover the source temperature: for the 60 K source, and
with $\sigma$=5\%, only $\sim$20\% of the fits recover temperatures
within 10\% of the source temperature; for the 100 K source, only
$\sim$10\% of the fits accurately recover the source temperature
within 10\%.

Also shown in Figure \ref{few} (as a dashed line) is the best fit
hyperbola to {\it PRONAOS} and {\it IRAS} data found by
\citet{Dupacetal03}.  The wavelengths of the five fluxes considered in
the fits in Figure \ref{few} are the same as those considered by
\citet{Dupacetal03}.  After describing the degeneracy between $\beta$
and $T$ in the next section, we discuss implications of our analysis
on recent observational interpretations in $\S$\ref{compsec}.

\section{Degeneracy between the Spectral Index and the Temperature}

We have shown that modest levels of noise can result in grossly
misleading $T$ and $\beta$ estimates from a least squares fit of
equation (\ref{fd}).\footnote{Other techniques, such as Bayesian or
likelihood based methods, may provide better estimates of the source
parameters.}  The best fit $\beta$ is anti-correlated with the best
fit $T$ due the degeneracy between $\beta$ and $T$ \citep{Dupacetal01,
Blainetal03, Sajinaetal06, Desertetal08}.

\begin{figure}
\plotone{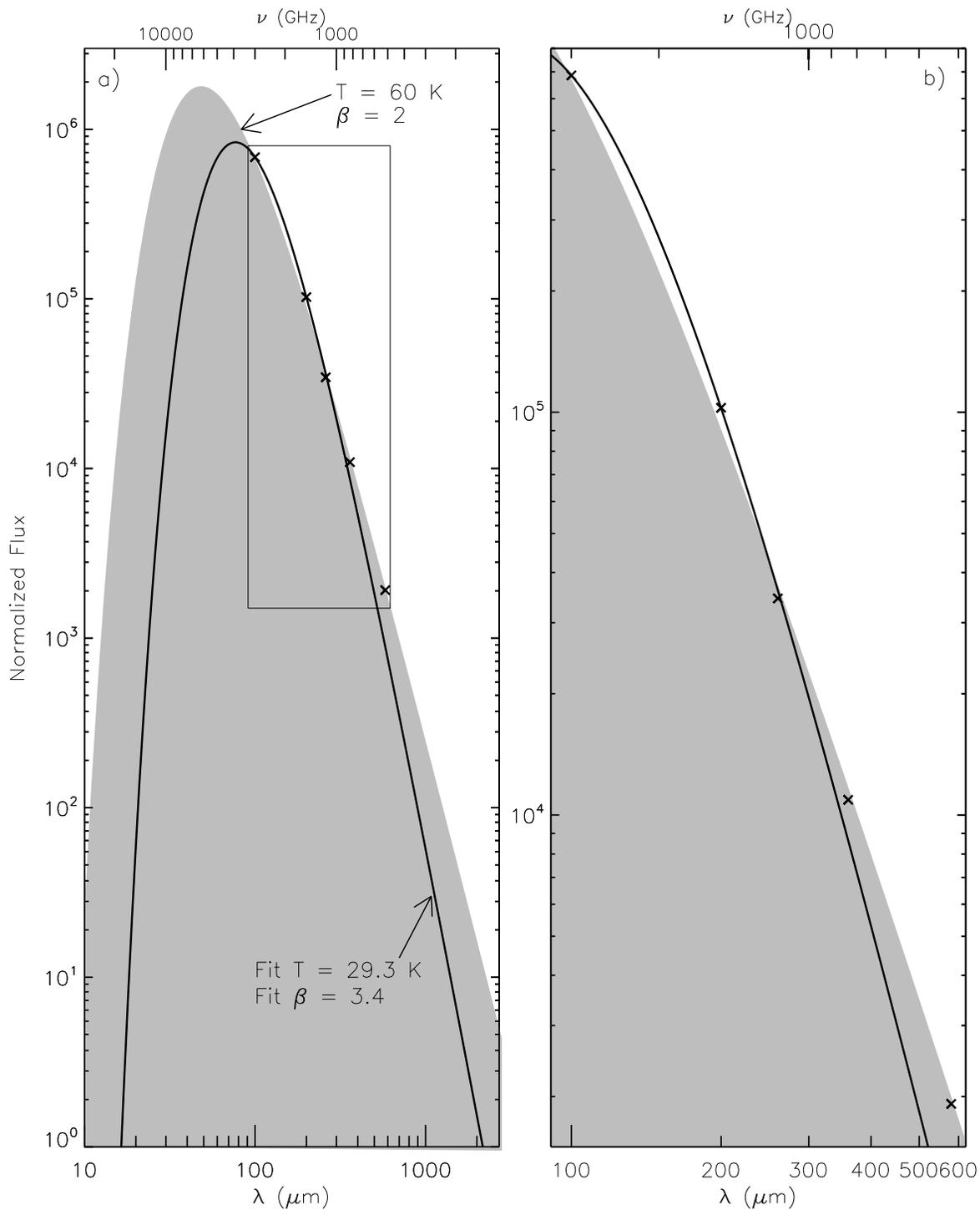}
\caption{Actual and fit SEDs from a 60 K isothermal source.  The
boundary of the shaded region is the dust SED from a 60 K source.  The
solid curve shows a fit to the few noisy fluxes (crosses) at
$\lambda$=100, 200, 260, 360, and 580 \micron.  The errors in those
fluxes are 2.1\%, 13.2\%, $-$7.5\%, $-$7.0\%, and $-$4.0\%
respectively.  Detail of boxed region in {\it (a)} is shown in {\it
(b)}.}
\label{sedfit}
\end{figure}

In the R-J part of the SED, fits to fluxes from various warm sources
may be difficult to distinguish (e.g. the 60 and 100 K fits in Figure
\ref{few}).  The reason for the confusion in the resulting fits is
illustrated in Figure \ref{sedfit}.  Figure \ref{sedfit}(a) compares
two SEDs, with $T$=60 K and 29.3 K, and $\beta$=2 and 3.4,
respectively.  The latter is a fit to five noisy fluxes from the 60 K
source.  The two SEDs are rather similar at large wavelengths,
especially in the range $\lambda \in$ 100-600 \micron, as shown in
Figure \ref{sedfit}(b).  In this wavelength range, there are numerous
combinations of $T$ and $\beta$ that are consistent with the observed
SED, and therefore it is difficult to constrain them.  Thus, fitting
modified blackbody SEDs solely to fluxes in the R-J regime produces
unreliable $T$ and $\beta$ estimates.  Moreover, the degeneracy
between $T$ and $\beta$ is such that an overestimate in $\beta$
results in an underestimate in $T$.

For isothermal sources, fits to fluxes near the peak of the SED
provide more accurate estimates of the source parameters (Paper I).
Thus, the 100-600 \micron\ wavelength range considered here can be
shown to provide accurate estimates for source with $T$\aplt20 K.
This is why the mean value of $T$ and $\beta$ recovers the source
values for the cold 20 K source in Figures \ref{range} and \ref{few},
but not so for the warmer sources.  Shorter wavelength observations in
the Wien regime may mitigate the effect of noise for warmer sources;
but, at wavelengths $\lambda$\aplt100 \micron, embedded sources as
well as transiently heated very small grains (in cold regions) may
contribute to the observed flux \citep{Li&Draine01}.

\section{Comparison with Recent Observations \label{compsec}}

\citet{Dupacetal03} find an inverse correlation between the best fit
$T$ and $\beta$ from {\it PRONAOS} data.  In their analysis, they
consider the degeneracy between $T$ and $\beta$ by performing
statistical tests on noisy model fluxes.  They find that noise is
insufficient to account for the derived $T-\beta$ trend, and suggest
that an inverse $T-\beta$ correlation is an intrinsic characteristic
of dust in the ISM.  In our extension of their work, we find a greater
scatter in $T$ and $\beta$ (see Fig. 4 of \citet{Dupacetal01}).  That
the best fits populate the $T$ and $\beta$ plane with a very similar
shape to that produced simply by the presence of noise is strongly
suggestive that noise is affecting these fitting results (see Figs
\ref{range} and \ref{few}).

The $T$ and $\beta$ fits from \citet{Dupacetal03} show lower values of
$\beta$ at $T$\apgt20 K, compared with the distribution in Figures
\ref{range} and \ref{few}.  As already discussed, comparing isothermal
sources with different values of $\beta$ would simply result in
vertically offset points on the $T-\beta$ plane.  We thus consider
whether sources with different temperatures, but with $\beta$=1.5, are
consistent with PRONAOS data, focusing on M17 \citep{Dupacetal02} and
Orion \citep{Dupacetal01}, which are the two sources that span the
full range $T\in$ 20-80 K in Figure 3 of \citet{Dupacetal03}.

\begin{figure}
\plotone{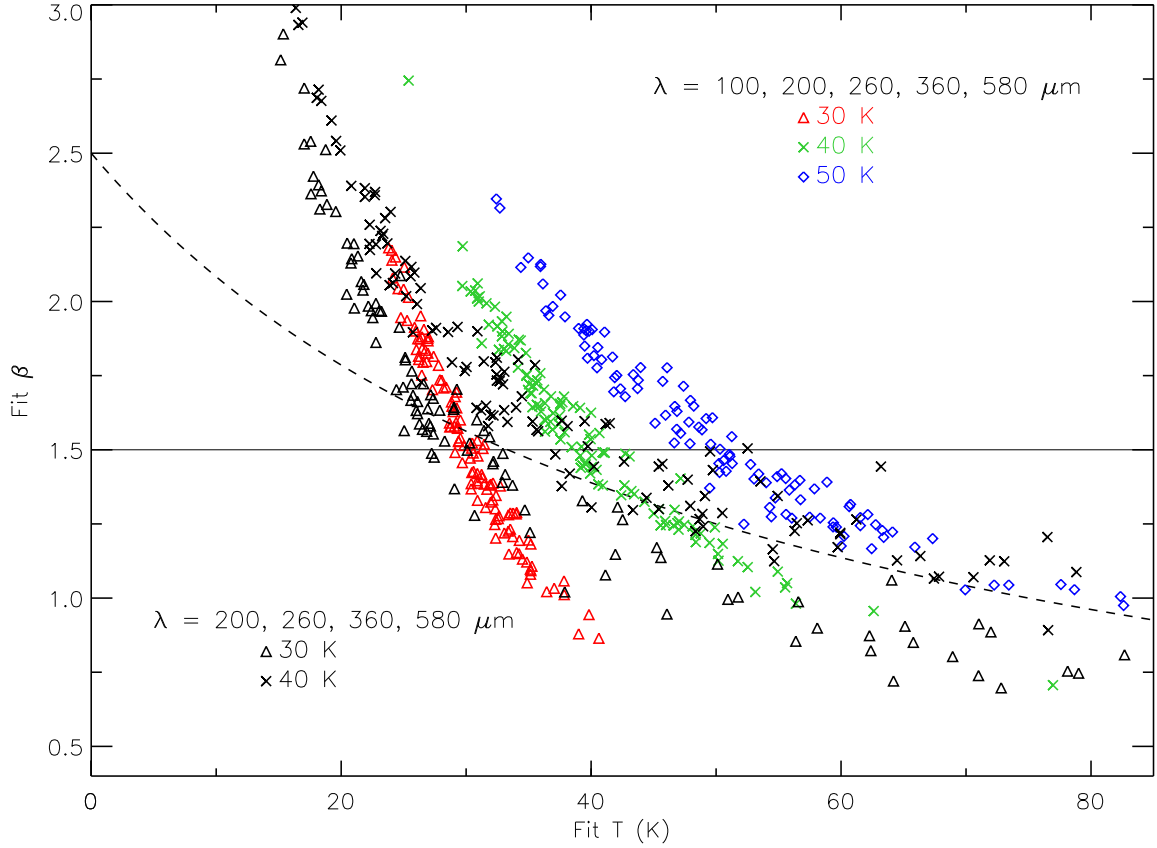}
\caption{Best fit $\beta$ and $T$ to noisy fluxes (with $\sigma$=5\%)
from isothermal sources with $T\in$30-50 K.  Colored points show fits
to fluxes with $\lambda$=100, 200, 260, 360, and 580 \micron.  Black
points show fits to fluxes from a 30 and a 40 K source excluding the
100 \micron\ flux.  The horizontal line indicates the spectral index
of the source, $\beta$=1.5.  The dashed line shows the best fit to
data presented by \citet{Dupacetal03}.}
\label{dupcomp}
\end{figure}

Figure \ref{dupcomp} shows numerous best fit $T$ and $\beta$ to five
noisy fluxes from 30 K, 40 K, and 50 K isothermal sources, with
$\beta$=1.5.  For the warmer sources, the fits are similar to many of
the points in Figure 4 of \citet{Dupacetal02} at T\apgt20 K.  As
\citet{Dupacetal02} describe, at some positions the IRAS 100 \micron\
fluxes were not included in the fit.  Figure \ref{dupcomp} also shows
such fits.  Though this simple model cannot account for the points at
low $T$ in Figure 4 of \citet{Dupacetal02}, many of the points at
$T$\apgt20 K are very similar.  Thus, a model with constant $\beta
\sim$ 1.5, for dust at temperatures \apgt30 K, may account for many of
the PRONAOS derived $T$ and $\beta$ points (at $T$\apgt20 K).  Such
temperatures may be realistic for much of the dust in M17
\citep[e.g.][]{Goldsmithetal97}.

\begin{figure}
\plotone{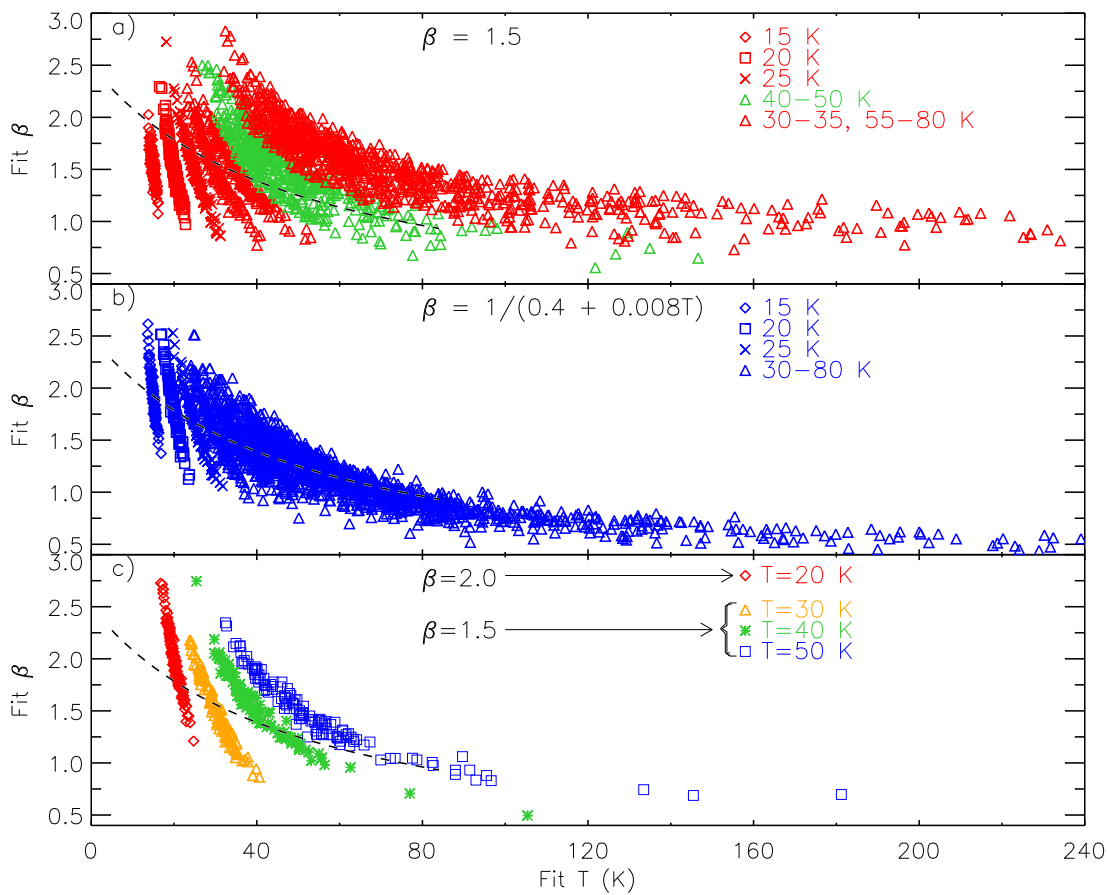}
\caption{Best fit $\beta$ and $T$ to noisy fluxes (with $\sigma$=5\%)
from isothermal sources with a range of temperatures, for three sets
of models, along with the best fit from \citet{Dupacetal03} (dashed
line).  (a) $\beta$=1.5; (b) $\beta$=1/(0.4+0.008$T$); and (c) a step
function in $\beta$, with $\beta$=1.5 for $T > 20$ K and $\beta$=2.0
for $T$=20 K.  The fluxes have $\lambda$=100, 200, 260, 360, and 580
\micron.}
\label{modscat}
\end{figure}

To verify whether any substantial differences can be easily identified
in cases where $\beta$ is constant or variable, we perform the
Monte-Carlo experiments described in $\S$\ref{sedfitsec} assuming
models with different $\beta-T$ dependencies.  The first model has
$\beta$=1.5 for all $T$; the second has $\beta$=1/(0.4+0.008$T$),
which is the fit found by \citet{Dupacetal03}.  We construct a series
of sources with $T \in$15-80 K, in increments of 5 K, and fit equation
(\ref{fd}) to numerous sets of noisy fluxes at 100, 200, 260, 360, and
580 \micron.  Figures \ref{modscat}(a)-(b) show the results for fits
to fluxes with noise levels of 5\%, for both models, along with the
best fit hyperbola from \citet{Dupacetal03}.

Figures \ref{modscat}(a)-(b) show that the fit parameters from the
lowest temperature sources at 15, 20, and 25 K are distinct from the
warmer sources.  As already discussed, at these temperatures, given
the wavelengths of the fluxes considered, the peaks of the emergent
SEDs are well sampled, so the fits are not very sensitive to noise.

The main difference between the constant-$\beta$ and the
inverse-hyperbolic $\beta(T)$ models is that the there is less scatter
in the fit $\beta$ from the model where $\beta$ depends on $T$.  This
difference occurs because the sources with low temperatures have
higher spectral indices (and vice versa); thus, this
inverse-hyperbolic $\beta(T)$ function reduces the scatter about
$\beta$ for a given $T$ relative to the scatter from the constant
$\beta$ case.

The extent of the scatter in Figure \ref{modscat} is dependent on the
chosen level of noise, the form of $\beta(T)$, as well as the {\it
temperatures of the sources considered in this test} (and the
wavelengths of the sampled fluxes).  For example, for the range
$T\in40-50$ K in the constant-$\beta$ model (green points in
Fig. \ref{modscat}(a)), the fits are well described by the best fit
hyperbola from \citet{Dupacetal03}.  Further, simply increasing the
noise level in the model where $\beta$ varies inversely with $T$ (Fig
\ref{modscat}(b)) could produce a $T-\beta$ scatter plot that is
similar to the constant $\beta$ model with the same temperature range
(Fig. \ref{modscat}(a)).  The primary effect of an intrinsic inverse
$T-\beta$ correlation is to reduce the spread in the $T-\beta$ plane
due to noise in the flux measurements.

Despite the differences in the scatter between the models in Figure
\ref{modscat}(a)-(b), there is a striking similarity in the shape of
the $T-\beta$ distribution: even though the models have different
forms in $\beta(T)$, fits to the fluxes result in a similar inverse
$T-\beta$ correlation.  Due to this similarity, it is very difficult
to distinguish an intrinsic $T-\beta$ anti-correlation from the
artificial anti-correlation arising due to noise.  Alternative
estimates of the actual range in source temperatures, such as
molecular line observations or radiative transfer modeling, would be
necessary to accurately account for the spread in $T$ and $\beta$ due
to the intrinsic degeneracy between the parameters in least squares
SED fitting.

One method to quantify the contribution of measurement uncertainties
to derived correlations is the use of correlation coefficients
\citep[e.g.][]{Kelly07}.  \citet{Dupacetal01} (in $\S$4.2) compare the
correlation coefficient of their best fits from Orion observations to
a model with no correlation between $T$ and $\beta$.  The model has
numerous uncorrelated $T$ and $\beta$ pairs, from which noisy fluxes
are constructed; fits to those noisy fluxes result in a correlation
coefficient between $T$ and $\beta$ of -0.4.  The discrepancy between
this value and the value from the observations, -0.92, indicates that
a uniform and uncorrelated distribution in $T$ and $\beta$ can be
ruled out \citep[see also][]{Yang&Phillips07}.  Such a test, however,
has not ruled out scenarios where $\beta$ is constant, the temperature
range is limited, or both.

The correlation coefficients from $T-\beta$ fits to noisy fluxes from
any isothermal model are \aplt -0.90, which are remarkably similar to
the value from the fits to Orion observations.  In a case where there
is a small range in $T$, such as the green points in Figure
\ref{modscat}(a) ($T \in$ 40-50 K), we compute a correlation
coefficient of -0.8, also similar to the value found from the
observations.  These statistical tests suggest that a constant
$\beta$, with a limited range in the source temperature, cannot be
ruled out in interpreting the {\it PRONAOS} Orion observations.

If the {\it PRONAOS} observations detected sources with $T$\aplt20 K,
and given the relative flux uncertainties of $\sim$5\%
\citep{Dupacetal01,Pajotetal06}, then the lack of points with
$T$\aplt20 {\it and} $\beta$\aplt1.2 is suggestive that $\beta$ for
such cold sources cannot be a constant value of 1.5.  One possibility
is that $\beta$=2 for the coldest sources, but $\beta$ = 1.5 for
sources with $T$\apgt30 K.  Similar to an inverse hyperbolic function
in $\beta(T)$, such a step function would result in a $T-\beta$
distribution with less scatter compared to a scenario where $\beta$ is
constant.  Figure \ref{modscat}(c) shows the fits from this step
function in $\beta(T)$.  This model illustrates that other functional
forms of $\beta(T)$ may produce a $T-\beta$ distribution similar to
that found by the {\it PRONAOS} observations.

The observationally inferred $T-\beta$ correlation is certainly
consistent with, e.g. an inverse intrinsic hyperbolic $\beta(T)$
function.  Yet, our analysis suggests that the intrinsic functional
form of $\beta(T)$ cannot be reliably derived from a fit to the $T$
and $\beta$ values obtained from a least-squares SED fit to the
observed fluxes.  Other functional forms of $\beta(T)$, in combination
with the noise levels, may also be consistent with the observations.
The agreement between the $T-\beta$ correlation coefficients from fits
to {\it PRONAOS} Orion observations and our isothermal, single $\beta$
models suggest that $T$ and $\beta$ may not be correlated.  Other
statistical tests are needed to quantify the $T-\beta$ degeneracy.

A more complete understanding of the distribution in the source $T$
and $\beta$ is necessary to accurately infer any physical $T-\beta$
correlation.  The scatter in the best fit parameters about the
hyperbola in the \citet{Dupacetal03} study is likely a consequence of
various factors.  Since the wavelengths of the observed fluxes are in
the R-J part of the emergent SED from warm sources, noise does
significantly bias the fitting.  Additionally, the dataset presented
in \citet{Dupacetal03} contains numerous sources, with different
spectral indices and temperatures, between sources and perhaps even
within individual sources; these issues may all contribute to the
scatter in the distribution of $T$ and $\beta$.

\section{Summary}

We have shown that a natural consequence of fitting modified blackbody
SEDs to flux measurements, with modest noise uncertainties, is an
inverse correlation between the temperature $T$ and spectral index
$\beta$.  Such an inverse correlation arises even for isothermal
sources with a constant $\beta$.  Least squares fits to fluxes from the
Rayleigh-Jeans regime of the emergent SED are very sensitive to noise
uncertainties.  Consequently, various isothermal sources with
different $T$s but identical $\beta$s may be indistinguishable through
simple SED fits.

We find that the spurious $T-\beta$ anti-correlation due to noise,
from numerous isothermal sources with a limited range in $T$ and
$\beta$, is similar to the observationally inferred anti-correlation
($\S$5).  A hyperbolic $T-\beta$ relation may indeed be a physical
characteristic of dust, but the observational results show
similarities with other explanations, such as a single $\beta$ for
isothermal sources with $T$\apgt30 K, or a step function where $\beta
\approx 2$ for $T$\aplt20 K and $\beta \approx 1.5$ for $T$\apgt30 K.
Thus, least squares SED fits to measured fluxes may not be able to
reveal any intrinsic correlation between $\beta$ and $T$.

\acknowledgements We are grateful to Brandon Kelly, Xavier Dupac, and
Jonathan Foster for comments on the draft and useful discussions.  We
thank the referee for comments that helped focus this paper.  We also
thank Peter Teuben for assistance using NEMO software \citep{Teuben95}
to carry out our analysis.  S. Schnee acknowledges support from the
OVRO, which is supported by the NSF through grant AST 05-40399.  R.S.,
J. K., and A. G. acknowledge support from the Harvard Initiative in
Innovative Computing, which hosts the Star-Formation Taste Tests
Community at which further details on these results can be found and
discussed (see http://www.cfa.harvard.edu/$\sim$agoodman/tastetests).

\end{document}